# Avalanche to streamer transition in particle simulations

Chao Li, Ute Ebert, W.J.M. Brok

**The avalanche to streamer transition is studied and illustrated in a particle model. The results are similar to those of fluid models. However, when super-particles are introduced, numerical artefacts become visible. This underscores the need of models that are hybrid in space.**

Streamers are growing filaments of weakly ionized non-stationary plasma produced by a sharp ionization front that propagates into non-ionized matter. Streamers are used in industrial applications such as lighting [1], ozone generation and gas and water purification, and they occur in natural processes such as lightning and transient luminous events in the upper atmosphere [2].

Streamers can emerge from ionization avalanches in free space when the self-induced field becomes comparable to the applied field; this was recently reinvestigated in the framework of a fluid model in [3]. [4] has doubted that fluid models could be applied to avalanches in noble gas in a very low field; and in [5] striations were found in this case. In nitrogen and in intermediate fields, particle and fluid models give essentially the same results for planar ionization fronts, while there are growing deviations at larger fields as discussed in [6].

Here we show how a streamer arises from few seed electrons and a consecutive avalanche. The system is pure nitrogen at standard temperature and pressure in a constant high field of 100 kV/cm. We assume that the electrodes are far away or don't exist as in many natural discharges. We use a particle model, while qualitatively similar results in a fluid model are illustrated in Figure 1 in [3].

The particle model includes the complete electron velocity and energy distribution as well as the discrete nature of particles. However, the required computation resources grow with the number of particles and eventually exceed the limits of any computer. This difficulty can be counteracted by using super-particles carrying the charge and the mass of many physical particles, but super-particles in turn create unphysical fluctuations and stochastic heating as we will demonstrate below.

The simulation starts from 100 pairs of electrons and ions at one point and follows the initial particles and their offspring up to time 0.36 ns. In Fig 1, the simulated streamer is presented at two different moments: t = 0.18 ns (first row), t = 0.36 ns with real particles (second row) and at the same time in a super-particle simulation (third row). The super-particles are introduced as follows: If a specified number of particles, in our case $10^5$, has been reached, a particle remapping is applied to reduce the number of computational particles; in this step, half of them are thrown away randomly and the weight of the remaining computational particles is doubled.

At time t = 0.18 ns, we have approximately $10^5$ electrons, and the maximal field enhancement is ~1%. The discharge is still in the avalanche phase. At t = 0.36 ns, the total number of electrons is roughly $1.5 \times 10^7$. Within the real particle simulation, the space charge layer has clearly formed. The maximal field enhancement is ~50%. In the super-particle simulation with $7 \times 10^4$ computational particles of mass 256, no charge layer has formed, but the charge densities are noisy and their maximum is inside the discharge. Clearly a different approach is needed for dealing computationally with the large numbers of electrons. In [7], we describe how to circumvent the unwieldy runtimes by coupling particle and fluid model in different spatial regions.

## References

[1] A. Bhoj and M. Kushner. *IEEE Trans on Plasmas Science*, **33**:518 (2005).
[2] V.P. Pasko, *Plasma Sources Sci. Technol.* **16**:13 (2007).
[3] C. Montijn and U. Ebert, *J. Phys. D: Appl. Phys.* **39**:2979 (2006).
[4] B.J.P. Dowds, R.K. Barrett and D.A. Diver, *Phys. Rev. E* **68**:026412 (2003).
[5] W.J.M. Brok, Ph.D. thesis, Eindhoven Univ. Tech., 2005, http://alexandria.tue.nl/extra2/200512799.pdf
[6] C. Li, W.J.M. Brok, U. Ebert, and J.J.A.M. van der Mullen. *J. Appl. Phys.*, **101**:123305 (2007).
[7] C. Li, U. Ebert, W.J.M. Brok and W. Hundsdorfer, *J.Phys. D: Appl. Phys. (Fast Track),* under revision.

Chao Li and Ute Ebert are with the *Center for Mathematics and Computer Science (CWI), Amsterdam, The Netherlands.*
W.J.M. Brok is with *Dept. Applied Physics, Eindhoven University of Technology, The Netherlands,* where U. Ebert also holds a part time employment.

The authors acknowledge support by the Dutch national program BSIK, in the ICT project BRICKS, theme MSV1.
Publisher Identifier S XXXX-XXXXXXXX-X

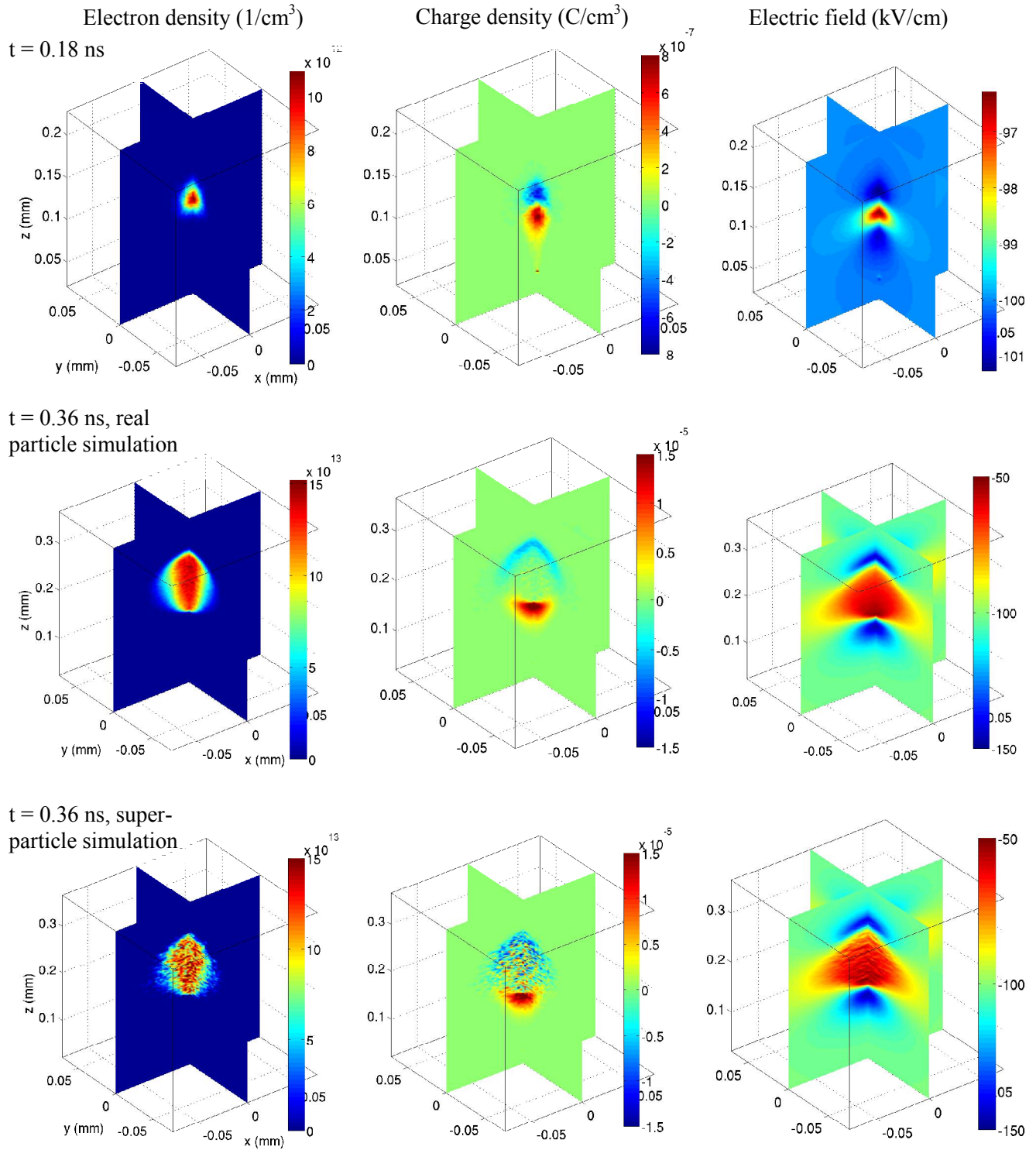

Fig. 1. Avalanche to streamer transition in a particle simulation in a field of 100 kV/cm in nitrogen at standard temperature and pressure. The first row shows the avalanche at t = 0.18 ns, the second row the streamer at t = 0.36 ns, and the third row the same as the second, but using super-particles that each represent 256 real particles. The columns show from left to right: electron density, charge density, and electric field.